\documentclass[11pt]{article}

\newcommand{\figdir}{figs-arXiv}
\newif\ifarxiv
\arxivtrue

\makeatletter
\usepackage[letterpaper,textwidth=6.5in,textheight=9in,top=1in,
  headheight=14pt,headsep=23pt,footskip=30pt]{geometry}
\usepackage[T1]{fontenc}
\usepackage{times}          %
\usepackage{fancyhdr}

\usepackage[square,comma]{natbib}
\usepackage[dvipsnames]{xcolor}
\usepackage{hyperref}
\definecolor{arxivlink}{RGB}{65,60,160}
\hypersetup{colorlinks=true,linkcolor=arxivlink,citecolor=arxivlink,
  urlcolor=arxivlink,breaklinks=true}

\newcommand{\runningtitle}{PrO Filtering with KMED}
\renewcommand{\headrulewidth}{0pt}
\fancypagestyle{plain}{\fancyhf{}\renewcommand{\headrulewidth}{0pt}%
  \fancyfoot[C]{\thepage}}

\renewcommand{\section}{\@startsection{section}{1}{\z@}%
  {-1.8ex \@plus -0.4ex \@minus -0.2ex}{1.0ex \@plus 0.2ex}%
  {\normalfont\large\bfseries}}
\renewcommand{\subsection}{\@startsection{subsection}{2}{\z@}%
  {-1.4ex \@plus -0.4ex \@minus -0.2ex}{0.6ex \@plus 0.2ex}%
  {\normalfont\normalsize\bfseries}}
\renewcommand{\paragraph}{\@startsection{paragraph}{4}{\z@}%
  {1.0ex \@plus 0.4ex \@minus 0.2ex}{-1em}%
  {\normalfont\normalsize\bfseries}}

\newcommand{\authorsep}{\end{tabular}\hfil\linebreak[0]\hfil
  \begin{tabular}[t]{c}\ignorespaces}
\renewcommand{\maketitle}{%
  \par
  \begingroup
  \thispagestyle{plain}%
  \setlength{\parskip}{0pt}%
  \centering
  \vspace*{\dimexpr 0.1in-\topskip\relax}%
  \hrule height 2pt          %
  \vspace{0.25in}%
  {\LARGE\scshape \@title\par}%
  \vspace{0.23in}%
  \hrule height 2pt          %
  \vspace{0.345in}%
  \let\And\authorsep
  \begin{tabular}[t]{c}\ignorespaces\@author\end{tabular}\par
  \vspace{0.2in}%
  \@date\par
  \vspace{0.12in}%
  \endgroup
  \let\maketitle\relax}

\renewenvironment{abstract}{%
  \centerline{\large\bfseries\scshape Abstract}%
  \vspace{0.5ex}%
  \begin{quote}}{\par\end{quote}\vskip 1ex}
\newcommand{\keywords}[1]{%
  \vskip 0.5ex\noindent\textbf{\textit{Keywords}}\enspace
  {\def\and{\unskip\ $\cdot$\ \ignorespaces}#1}\par\vskip 1ex}
\makeatother

\usepackage{microtype}
\usepackage{graphicx}
\usepackage{tabularx}
\usepackage{subcaption}
\usepackage{booktabs}
\usepackage{float}
\usepackage{amsmath}
\usepackage{amssymb}
\usepackage{mathtools}
\usepackage{amsthm}
\usepackage[capitalize,noabbrev]{cleveref}
\usepackage[font=small,skip=4pt]{caption}
\usepackage{enumitem}

\setlist{itemsep=2pt,topsep=2pt,parsep=0pt,partopsep=0pt}
\makeatletter
\g@addto@macro\normalsize{%
  \setlength\abovedisplayskip{6pt plus 2pt minus 3pt}%
  \setlength\belowdisplayskip{6pt plus 2pt minus 3pt}%
  \setlength\abovedisplayshortskip{2pt plus 2pt}%
  \setlength\belowdisplayshortskip{4pt plus 2pt minus 2pt}}
\makeatother
\usepackage{aliascnt}
\theoremstyle{plain}

\newaliascnt{proposition}{theorem}\newtheorem{proposition}[proposition]{Proposition}\aliascntresetthe{proposition}
\newaliascnt{lemma}{theorem}\aliascntresetthe{lemma}
\newaliascnt{corollary}{theorem}\aliascntresetthe{corollary}
\theoremstyle{definition}
\newaliascnt{definition}{theorem}\aliascntresetthe{definition}
\newaliascnt{assumption}{theorem}\aliascntresetthe{assumption}
\theoremstyle{remark}
\newaliascnt{remark}{theorem}\newtheorem{remark}[remark]{Remark}\aliascntresetthe{remark}

\input{preamble}

\title{PrO Filtering via Kernel Mean Embedding Dynamics}

\author{%
  Linfeng Wang \\
  King's College London, UK \\
  \texttt{linfeng.2.wang@kcl.ac.uk}
  \And
  Nikolas N\"{u}sken \\
  King's College London, UK\\
  \texttt{nikolas.nusken@kcl.ac.uk}
  \And
  Chris. J. Oates \\
  Newcastle University, UK\\
  \texttt{chris.oates@ncl.ac.uk}
}

\date{\today}

\hypersetup{
  pdftitle={PrO Filtering via Kernel Mean Embedding Dynamics},
  pdfauthor={Linfeng Wang, Nikolas N\"usken, Chris. J. Oates},
  pdfkeywords={kernel methods, measure transport, post-Bayesian, predictively-oriented posterior},
}

\begin{document}
\maketitle

\begin{abstract}
Filtering methods are routinely deployed with misspecified state-space models, where Bayesian filters can diverge and become overconfident.
Predictively-oriented (PrO) filtering is a recent, hyperparameter-free alternative that was demonstrated to outperform (generalised) Bayesian filtering methods when either the dynamical model or the measurement model is structurally misspecified.
This can be attributed to \emph{hedging} behaviour: when an observation conflicts with the forecast, the PrO update places mass near both.
Yet existing implementations of PrO filtering rely on unimodal (Gaussian) approximations, whose ability to effectively hedge is limited.
This paper develops the first nonparametric, particle-based implementation of PrO filtering.
The key obstacle, that PrO filtering distributions are only implicitly defined, is overcome by constructing a path from the predictive to the PrO filtering distribution whose rate is available in closed form, enabling the use of kernel mean embedding dynamics for measure transport.
The resulting algorithm requires only simulation from the dynamical model and evaluations of the likelihood, with neither importance weights nor the predictive density. Moreover, as we prove, the number of pseudo-time steps it requires does not grow with the outlyingness of the observation, in contrast to transport along the Bayesian tempering path.
An empirical assessment shows substantial improvement over Gaussian variation of PrO filtering in challenging nonlinear examples where Gaussian approximation is unreasonably limited.

\end{abstract}

\keywords{Kernel methods \and Measure transport \and Post-Bayesian \and Predictively-oriented posterior}

\section{Introduction}
\label{sec:intro}

Filtering, the sequential inference of the latent state of a dynamical system from noisy observations, is central to time series analysis \citep{durbin2012time}, target tracking \citep{chen2011kalman}, robotics \citep{thrun2002probabilistic} and data assimilation \citep{Evensen2009}.
Almost all filtering methods target the Bayesian filtering distribution, and thus implicitly trust the state-space model to be correct.
In practice it never is, and under misspecification Bayesian filters can become overconfident and even diverge \citep{schlee1967divergencekf,fitzgerald2003divergencekf}; robust alternatives typically target a specific form of misspecification or introduce additional hyperparameters to expand the capacity of the state-space model (cf. \Cref{sec:related}).

\begin{figure*}[t]
  \centering
  \ifarxiv
  \includegraphics[width=0.95\textwidth]{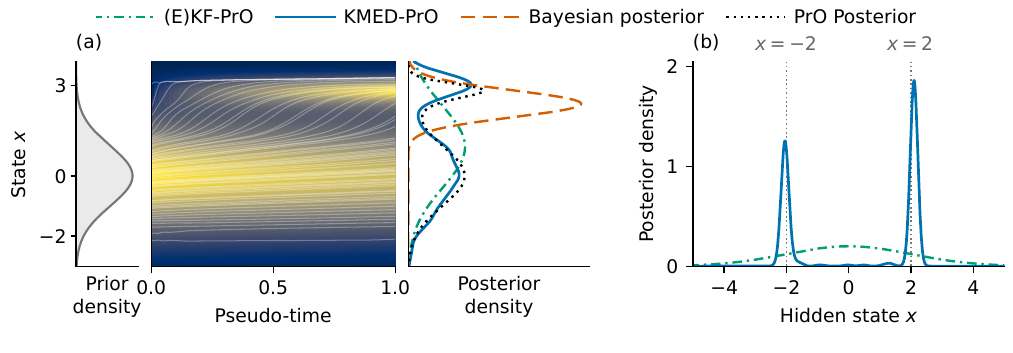}
  \else
  \includegraphics[width=\textwidth]{figs-arXiv/intro_limitations.pdf}
  \fi
    \caption{Illustration of the recovery of the PrO posterior via KMED. Experimental details are given in Appendix~\ref{app:scalar-experiments}. (a) KMED-PrO yields a more accurate representation of the PrO posterior than the Gaussian approximation (EKF-PrO). (b) EKF-PrO can become trapped when the Jacobian of the measurement model vanishes, whereas KMED-PrO provides a good particle-based approximation.}
  \label{fig:teaser}
  \label{fig:static}
\end{figure*}

\emph{Predictively oriented} (PrO) filtering \citep{shen2026predictively} offers a principled, hyperparameter-free alternative.
By replacing the expected log-likelihood in the variational characterisation of Bayes' rule with the log of the predictive likelihood, PrO filtering selects the filtering distribution whose \emph{predictions} best explain the data.
When an observation conflicts with the forecast, the PrO update \emph{hedges}, retaining mass near the forecast while placing mass near the observation (\Cref{fig:teaser}(a)).
As such, PrO filtering has been shown to outperform Bayesian and generalised Bayesian filters under structural misspecification of either the dynamical or the measurement model \citep[Section 3]{shen2026predictively}.

However, PrO filtering distributions are defined only implicitly, as the solution of a self-consistent equation, and all existing implementations restrict them to be Gaussian: the Kalman and extended Kalman filter (EKF) versions, KF-PrO and EKF-PrO, of \citet{shen2026predictively}, and the unscented version, UKF-PrO, of \citet{kean2026unscented}.
This restriction is severe, because a Gaussian cannot easily hedge: the exact PrO update can be bimodal, whereas the Gaussian approximation places a single inflated mode between the forecast and the observation (\Cref{fig:teaser}(a)), so that the full potential of PrO filtering is not realised.
Gaussian implementations also rely on linearisation or sigma points, a further source of error in the nonlinear settings where misspecification is often most pronounced.
For example, when the Jacobian of the measurement model vanishes at the current mean, EKF-PrO receives no information from the data and remains at the prior (\Cref{fig:teaser}(b)).

This observation motivates seeking instead a flexible, particle-based \ac{PrO} filtering method.
However, developing a particle-based implementation is not entirely straightforward.
Reweighting the predictive ensemble cannot place mass near an observation that lies far from the current ensemble, which is the situation where hedging is most needed; score-based methods, including those described in \citet{shenprediction,chazal2025computable,liu2025detecting,mclatchie2026markov}, require the score of the predictive distribution, which is not available in a particle-based filtering context; and transport-based filters \citep{daum2010exact,reich2011dynamical,wang2024measure} are built around the likelihood-tempering path of Bayes' rule, which has no PrO counterpart. These limitations are summarised in \Cref{tab:comparison}.

\paragraph{Our Contributions.}
This paper develops the first nonparametric, particle-based implementation of PrO filtering.
Our contributions are:
\begin{itemize}
	\item \textbf{A transport path for PrO.}
	We introduce a novel homotopy for updating the PrO objective which admits a closed-form rate (Proposition \ref{prop:rate}), even though the PrO filtering distribution is only implicitly defined.
	\item \textbf{A derivative-free, weight-free algorithm.}
	Combining this path with the kernel mean embedding dynamics (KMED) of \citet{wang2024measure, maurais2024sampling} yields KMED-PrO (\Cref{alg:kmed_pro}), which transports an equally weighted particle ensemble from the predictive to the PrO filtering distribution (\Cref{fig:teaser}(a)). It is derivative-free: it requires only simulation from the dynamical model and pointwise likelihood evaluations, with no model Jacobians or likelihood gradients; it also needs no importance weights or predictive density, and can represent multimodal filtering distributions.
	\item \textbf{Cost that does not grow with outlyingness.}
	The PrO path has Fisher--Rao length at most one, uniformly over the observation (Proposition \ref{prop:length}), whereas the length of the Bayesian path can be arbitrarily large; we leverage this fact to design an adaptive discretisation scheme whose number of pseudo-time steps is bounded independently of how surprising the observation is.
	\item \textbf{Empirical gains.}
	On a partially observed Langevin diffusion, KMED-PrO retains distinct state branches under ambiguous observations, and in range tracking with an omitted persistent force it improves predictive scores over EKF-PrO and the Bayesian EKF (EKF-Bayes).
\end{itemize}

\newcommand{\comparisonCaption}{\textbf{Comparison of filtering methods.}
	Columns: (a) targets PrO; (b) non-Gaussian; (c) derivative-free; (d) needs the predictive distribution only through samples; (e) unweighted particles, i.e.\ no importance weights or resampling.
	$^\dagger$EKF-PrO requires Jacobians of the model; KF-PrO applies only to linear models.
	$^\ddagger$Developed for PrO posteriors in the static setting.}
\newcommand{\comparisonTabular}{%
	\begin{tabular}{lccccc}
		\toprule
		Method & (a) & (b) & (c) & (d) & (e) \\
		\midrule
		Particle filter & $\times$ & $\checkmark$ & $\checkmark$ & $\checkmark$ & $\times$ \\
		EnKF & $\times$ & $\times$ & $\checkmark$ & $\checkmark$ & $\checkmark$ \\
		KMED \citep{wang2024measure} & $\times$ & $\checkmark$ & $\checkmark$ & $\checkmark$ & $\checkmark$ \\
		(E)KF-PrO \citep{shen2026predictively} & $\checkmark$ & $\times$ & $\times^\dagger$ & $\times$ & $\checkmark$ \\
		UKF-PrO \citep{kean2026unscented} & $\checkmark$ & $\times$ & $\checkmark$ & $\times$ & $\checkmark$ \\
		VGD \citep{chazal2025computable}$^\ddagger$ & $\checkmark$ & $\checkmark$ & $\times$ & $\times$ & $\checkmark$ \\
		\midrule
		KMED-PrO (ours) & $\checkmark$ & $\checkmark$ & $\checkmark$ & $\checkmark$ & $\checkmark$ \\
		\bottomrule
	\end{tabular}}
\ifarxiv
\begin{table}[t]
	\small
	\begin{minipage}[c]{0.55\textwidth}
	\centering
	\setlength{\tabcolsep}{5pt}
	\comparisonTabular
	\end{minipage}\hfill
	\begin{minipage}[c]{0.42\textwidth}
	\caption{\comparisonCaption}
	\label{tab:comparison}
	\end{minipage}
\end{table}
\else
\begin{table}[t]
	\caption{\comparisonCaption}
	\label{tab:comparison}
	\vskip 0.1in
	\centering
	\small
	\setlength{\tabcolsep}{5pt}
	\comparisonTabular
\end{table}
\fi

\paragraph{Outline.}
The remainder of the manuscript is structured as follows:
\Cref{sec:methods} introduces the mathematical set-up and then sets out our methodology; \Cref{sec:experiments} contains a detailed empirical assessment; \Cref{sec:related} relates our contribution to earlier work; and a brief discussion concludes in \Cref{sec:discussion}.

\section{Methodology}
\label{sec:methods}

The problem setting is described in \Cref{subsec:setting}, and PrO filtering is recalled in \Cref{subsec:filtering_as_opt}.
A homotopy from the predictive to the PrO filtering distribution is constructed in \Cref{subsec:homotopy}, and KMED is generalised to operate on this novel homotopy in \Cref{subsec:kmed}.
Practical considerations are discussed in \Cref{subsec:practical}.

\subsection{Problem Setting}
\label{subsec:setting}

We consider a state-space model
\begin{align}
	x_n & = \mathcal{M}(x_{n-1}, \epsilon_n) , \label{eq:dynamics} \\
	y_n & = \mathcal{O}(x_n, \xi_n) , \label{eq:measurement}
\end{align}
where $x_n \in \R^d$ is a latent state indexed by discrete time $n \in \{1,2,\dots\}$, $y_n$ is an observation, and $\epsilon_n$ and $\xi_n$ are independent noise processes.
The dynamical model \eqref{eq:dynamics} and the measurement model \eqref{eq:measurement} induce, respectively, a transition density $p(x_n | x_{n-1})$ and a likelihood $p(y_n | x_n)$.
Only the $y_n$ are observed, and the task of \emph{filtering} is to sequentially infer the current state $x_n$ from the current observation set $\{y_1, \dots, y_n\}$.
Additional static parameters $\theta$ can be incorporated into \eqref{eq:dynamics} and \eqref{eq:measurement} by augmenting the state vector $x$ with an additional component $\theta$ that remains static under the dynamical model; as such we do not make any distinction in the sequel.

\subsection{Filtering as an Optimisation Task}
\label{subsec:filtering_as_opt}

This section recalls, from \citet{shen2026predictively}, how probabilistic filtering can be cast as an optimisation task.
Let $\mathcal{P}(\R^d)$ denote the set of probability distributions on $\R^d$, which we identify with their densities where these exist.
Given a filtering distribution $q_{n-1|n-1}$ at time $n-1$, the \emph{prediction step} computes
\begin{align}
	q_{n|n-1}(x_n) = \int p(x_n | x_{n-1}) \, q_{n-1|n-1}(x_{n-1}) \, \dd x_{n-1} , \label{eq:prediction}
\end{align}
and the \emph{update step} computes
\begin{align}
	q_{n|n} = \argmin_{q \in \mathcal{P}(\R^d)} \; \ell(y_n ; q) + \KL(q \,\|\, q_{n|n-1}) , \label{eq:update}
\end{align}
for some loss function $\ell$.
Bayesian filtering corresponds to the average negative log-likelihood
\[
  \ell^{\mathrm{Bayes}}(y_n;q) = - \int \log p(y_n | x_n) \, q(x_n) \, \dd x_n,
\]
for which \eqref{eq:update} is solved by $q_{n|n}(x) \propto q_{n|n-1}(x) \, p(y_n | x)$, recovering Bayes' rule.
PrO filtering instead uses
\begin{align}
	\ell^{\mathrm{PrO}}(y_n ; q) = - \log p_q(y_n) , \label{eq:pro_loss}
\end{align}
where $p_q(y_n) := \int p(y_n | x_n) \, q(x_n) \, \dd x_n$.
This is the logarithmic score \citep{gneiting2007strictly} of the predictive distribution $p_q$ induced by $q$.
The expectation of \eqref{eq:pro_loss} with respect to the true distribution of $y_n$ is, up to an additive constant, the Kullback--Leibler divergence from the true distribution to $p_q$; the PrO loss therefore rewards filtering distributions  $q$ whose predictions are accurate, rather than filtering distributions that concentrate on the most likely state.
Moreover, since $q \mapsto p_q(y_n)$ is linear and $-\log$ is convex, the objective in \eqref{eq:update} is strictly convex in $q$ when $\ell = \ell^{\mathrm{PrO}}$, so that the PrO filtering distribution is unique whenever it exists.

Setting the first variation of the objective in \eqref{eq:update} to a constant shows that the PrO filtering distribution satisfies
\begin{align}
	q_{n|n}(x) \propto q_{n|n-1}(x) \exp\left( \frac{p(y_n | x)}{p_{q_{n|n}}(y_n)} \right) . \label{eq:pro_fixed_point}
\end{align}
This equation is only an implicit characterisation, since $q_{n|n}$ appears on both sides.
In addition, \eqref{eq:pro_fixed_point} reveals that the PrO filtering distribution is non-Gaussian even in the linear--Gaussian setting, since the exponent in \eqref{eq:pro_fixed_point} is a Gaussian density in $x$, rather than a quadratic form.
To see that this departure from Gaussianity can be substantial, consider ${d = 1}$, $q_{n|n-1} = \mathcal{N}(0,1)$ and $p(y_n | x) = \mathcal{N}(y_n ; x, 0.5^2)$, with an observation $y_n = 3$ that is surprising under the predictive distribution.
Solving \eqref{eq:pro_fixed_point} numerically yields a bimodal $q_{n|n}$ (\Cref{fig:teaser}(a)), with one mode at the predictive mean and a second mode close to the observation, the latter carrying roughly one third of the probability mass.
In other words, the PrO update \emph{hedges} between the predictive distribution and the observation, in a manner that reflects their mutual inconsistency.
\citet{shen2026predictively} restricted \eqref{eq:update} to the set of Gaussian distributions, so that the hedging is necessarily replaced by a single, inflated mode (\Cref{fig:teaser}(a)); this is a poor approximation to the \ac{PrO} filtering distribution and motivates the nonparametric approach that we develop next.

\subsection{A Homotopy for the PrO Update}
\label{subsec:homotopy}

Throughout this section, the time index $n$ is fixed and suppressed where possible.
Transport methods such as KMED need a path $(\pi_t)_{t \in [0,1]}$ from $\pi_0 = q_{n|n-1}$ to $\pi_1 = q_{n|n}$.
For Bayesian filtering, \citet{wang2024measure} take
\begin{align}
	\pi_t = \frac{e^{-th} \pi_0}{Z_t} , \qquad h(x) = - \log p(y_n | x) , \label{eq:bayes_path}
\end{align}
with $Z_t = \int e^{-th} \, \dd \pi_0$, which satisfies $\partial_t \log \pi_t = -(h - \E_{\pi_t}[h])$. To generalise this approach to a wider class of variational objectives, including \ac{PrO} filtering, we consider the homotopy 
\begin{align}
	\pi_t = \argmin_{q \in \mathcal{P}(\R^d)} \; t \, \ell^{\mathrm{PrO}}(y_n ; q) + \KL(q \,\|\, \pi_0) , \label{eq:pro_path}
\end{align}
for $t \in [0,1]$.
Each $\pi_t$ is unique by strict convexity, and clearly $\pi_0 = q_{n|n-1}$ and $\pi_1 = q_{n|n}$.
Analogously to \eqref{eq:pro_fixed_point}, each $\pi_t$ satisfies $\pi_t \propto \pi_0 \exp(t \, r_t)$, where
\begin{align}
	r_t(x) := \frac{p(y_n | x)}{p_{\pi_t}(y_n)} \label{eq:ratio}
\end{align}
is the ratio of the likelihood to the predictive likelihood under $\pi_t$.
Note that $\E_{\pi_t}[r_t] = 1$, and let $\sigma_t^2 := \Var_{\pi_t}[r_t]$.
In contrast to \eqref{eq:bayes_path}, the PrO path is not a tempering path, since the exponent $t \, r_t$ depends on $\pi_t$ itself.
Nevertheless, its rate can be computed in closed form:

\begin{proposition}[Rate of the PrO path]
	\label{prop:rate}
	Assume that $\sup_x p(y_n | x) < \infty$ and that $t \mapsto \pi_t$ is differentiable, in the sense that $\partial_t \log \pi_t$ exists and differentiation can be interchanged with integration against $\pi_t$.
	Then
	\ifarxiv
	\begin{align}
		\partial_t \log \pi_t = -(h_t - \E_{\pi_t}[h_t]) , \qquad h_t(x) := - \frac{r_t(x) - 1}{1 + t \sigma_t^2} . \label{eq:pro_rate}
	\end{align}
	\else
	\begin{align}
		\hspace{-5pt} \partial_t \log \pi_t = -(h_t - \E_{\pi_t}[h_t]) , \; h_t(x) := - \frac{r_t(x) - 1}{1 + t \sigma_t^2} . \label{eq:pro_rate}
	\end{align}
	\fi
	Moreover, the predictive likelihood evolves according to $\partial_t \log p_{\pi_t}(y_n) = \sigma_t^2 / (1 + t \sigma_t^2)$.
\end{proposition}

\noindent
The proof is contained in \Cref{app:proof_rate}.
Since $\E_{\pi_t}[r_t] = 1$ we have $\E_{\pi_t}[h_t] = 0$, but we write \eqref{eq:pro_rate} in a form that emphasises the analogy with the Bayesian path \eqref{eq:bayes_path}: the PrO path is obtained by replacing the static potential $h$ with the time-dependent potential $h_t$, which depends on $\pi_t$ through $r_t$ and $\sigma_t^2$.

\begin{remark}[Comparison with Bayes]
	\label{rem:bayes_vs_pro}
	The PrO potential $h_t$ differs from its Bayesian counterpart $h$ in two respects.
	First, the log-likelihood is replaced by the likelihood ratio $r_t$, which is invariant to the scale of the likelihood and, unlike the log-likelihood, is bounded below.
	Second, the factor $(1 + t \sigma_t^2)^{-1}$ accounts for the fact that the predictive likelihood $p_{\pi_t}(y_n)$ in \eqref{eq:ratio} moves as $\pi_t$ moves.
	The correction factor damps the update when $\sigma_t^2$ is large, which occurs when the likelihood varies substantially across the support of $\pi_t$; this can be viewed as a nonparametric analogue of the conservativity of KF-PrO relative to the Kalman filter established in \citet{shen2026predictively}.
\end{remark}

A striking consequence of Proposition \ref{prop:rate} is that the PrO path is always short:

\begin{proposition}[PrO paths are short]
	\label{prop:length}
	Under the assumptions of Proposition \ref{prop:rate},
	\begin{align}
		\| \partial_t \log \pi_t \|_{L^2(\pi_t)} = \frac{\sigma_t}{1 + t \sigma_t^2} \leq \min\left\{ \sigma_t , \frac{1}{2 \sqrt{t}} \right\} . \label{eq:speed_bound}
	\end{align}
	Consequently the Fisher--Rao length of the PrO path satisfies $\int_0^1 \| \partial_t \log \pi_t \|_{L^2(\pi_t)} \, \dd t \leq 1$.
	Moreover, for each $t \in [0,1]$, the symmetrised KL divergence satisfies
	\begin{align}
	& \mathrm{SKL}(\pi_t \, \| \, \pi_0) \label{eq:sym_kl} \\
		& := \KL(\pi_t \,\|\, \pi_0) + \KL(\pi_0 \,\|\, \pi_t) = t \left( 1 - \frac{p_{\pi_0}(y_n)}{p_{\pi_t}(y_n)} \right) \leq t , \nonumber
	\end{align}
	and thus $\mathrm{SKL}(q_{n|n} \,\|\, q_{n|n-1}) \leq 1$.
\end{proposition}

\noindent
The proof is contained in \Cref{app:proof_length}.
The bounds in Proposition \ref{prop:length} hold uniformly over the observation $y_n$, the predictive distribution and the likelihood. 
For the Bayesian path \eqref{eq:bayes_path}, by contrast, $\| \partial_t \log \pi_t \|_{L^2(\pi_t)}$ is the standard deviation of the log-likelihood under $\pi_t$, which can be arbitrarily large when $y_n$ is outlying.
Likewise, no analogue of \eqref{eq:sym_kl} holds for Bayesian filtering: if $q_{n|n-1} = \mathcal{N}(0,1)$ and $p(y_n | x) = \mathcal{N}(y_n; x, 1)$, then the Bayesian filtering distribution is $\mathcal{N}(y_n/2, 1/2)$, and its symmetrised Kullback--Leibler divergence from $q_{n|n-1}$ is $1/4 + 3 y_n^2 / 8$, which grows without bound as $|y_n| \to \infty$.
Proposition \ref{prop:length} therefore provides a quantitative sense in which the PrO update is conservative, and in \Cref{subsec:practical} it will be used to control the number of pseudo-time steps required by our method.

\subsection{KMED for PrO Filtering}
\label{subsec:kmed}

Now we will generalise the KMED of \citet{wang2024measure, maurais2024sampling} to the homotopy \eqref{eq:pro_path};
recall that our aim is to construct an interacting particle system evolving in artificial time $t \in [0, 1]$ that connects the prediction $\pi_0 = q_{n | n-1}$ to the PrO update $\pi_1 = q_{n | n}$ through its empirical laws. As before, $n$ is fixed and suppressed in the notation.

Let $k : \R^d \times \R^d \to \R$ be a continuously differentiable positive definite kernel with \ac{rkhs} $\Hk$, canonical feature map ${\phi_k(x)=k(\cdot,x)}$ and mean embedding ${\Phi_k(\rho)=\int \phi_k(x)\,\rho(\dd x)}$.
We assume that the expectations and derivatives below exist.
For a characteristic kernel, $\Phi_k$ identifies the distribution uniquely \citep{muandet2017kernel}.
Consider the dynamics
\begin{equation}
  \frac{\dd X_t}{\dd t}=v_t(X_t), \qquad X_0\sim\pi_0,
  \label{eq:mean_field}
\end{equation}
with $\rho_t=\mathrm{Law}(X_t)$. In what follows, we construct the driving vector field $v_t$ in such a way that $\rho_t$ approximately reproduces the homotopy defined in \eqref{eq:pro_path}: $\rho_t \approx \pi_t$ for all $t \in [0, 1]$.

The kernel mean embedding of $\rho_t$ evolves as $\partial_t\Phi_k(\rho_t)=\E_{\rho_t}[\nabla_x\phi_k(x)\cdot v_t(x)]$, whereas Proposition \ref{prop:rate} gives $\partial_t\Phi_k(\pi_t)=-\Cov_{\pi_t}[\phi_k,h_t]$ for the prescribed homotopy. Equating these rates, and replacing $\pi_t$ by $\rho_t$ in both the expectation and the potential, yields
\begin{equation}
  -\int\nabla_xk(\cdot,x)\cdot v_t(x)\,\rho_t(\dd x)=f_t,
  \label{eq:kce}
\end{equation}
where $f_t=\Cov_{\rho_t}[\phi_k,h_t]$ and $h_t$ is defined by \eqref{eq:pro_rate} with $r_t=p(y_n|\cdot)/p_{\rho_t}(y_n)$ and $\sigma_t^2=\Var_{\rho_t}[r_t]$.
As in \citet{wang2024measure}, we seek a velocity of the form
\begin{equation}
  v_t(x)=-C_t\int\nabla_x k(x,y)\,\alpha_t(y)\,\rho_t(\dd y),
  \label{eq:ansatz}
\end{equation}
where $C_t\in\R^{d\times d}$ is a symmetric positive definite preconditioning matrix and ${\alpha_t:\R^d\to\R}$ is an unknown coefficient function.
Substitution into \eqref{eq:kce}, with Tikhonov regularisation $\varepsilon>0$, gives
\begin{subequations}
\label{eq:kmed}
\begin{align}
  v_t(X_t)&=-C_t\int\alpha_t(y)\nabla_{X_t}k(X_t,y)\,\rho_t(\dd y),
  \label{eq:kmed_ode}\\
  f_t&=\int G_{\rho_t,C_t}(\cdot,y)\alpha_t(y)\,\rho_t(\dd y)
       +\varepsilon\alpha_t,
  \label{eq:kmed_integral}
\end{align}
\end{subequations}
where $G_{\rho,C}(x,y)=\int\nabla_z k(x,z)\cdot C\nabla_z k(z,y)\,\rho(\dd z)$.
Notice that \eqref{eq:kmed} is a coupled system: \eqref{eq:kmed_integral} determines $\alpha_t$ via a linear equation, given the current distribution $\rho_t$; inserting $\alpha_t$ into \eqref{eq:kmed_ode} then determines the infinitesimal update of $X_t$ and its distribution $\rho_t$.
The PrO construction replaces the static Bayesian potential in KMED from \citet{wang2024measure, maurais2024sampling} with a potential that depends on both time and the evolving law.
The predictive distribution enters only through $X_0\sim\pi_0$: neither its density nor its score is required, unlike score-based transports such as Stein variational gradient descent \citep{liu2016stein}.

\begin{\algoenv}[t]
  \caption{KMED-PrO (one filtering step)}
  \label{alg:kmed_pro}
  \KwIn{posterior particles $\{x^i_{n-1}\}_{i=1}^N$ approximating $q_{n-1|n-1}$; observation $y_n$; kernel $k$; regularisation $\varepsilon>0$; step control $\delta>0$}
  Forecast $x^i_{n|n-1}=\mathcal M(x^i_{n-1},\epsilon_n^i)$ independently for $i=1,\ldots,N$\;
  Initialise $X^i_0 \leftarrow x^i_{n|n-1}$ for $i=1,\ldots,N$, and $t\leftarrow0$\;
  \While{$t<1$}{
    Compute $\hat r_t^i$, $\hat\sigma_t^2$, $\hat h_t^i$ using \eqref{eq:potential_particles}, and $C_t$ using \eqref{eq:empirical-covariance-preconditioning}\;
    Choose $\Delta t$ using \eqref{eq:adaptive_step}\;
    Compute $\mathbf G_t$ and $\boldsymbol h_t^k$ using \eqref{eq:definition-linear-algebra}\;
    Solve \eqref{eq:kmed_particles_linear} for $\boldsymbol\alpha_t$\;
    Update $X^i_{t+\Delta t} \leftarrow X^i_t + \Delta t\,v_t(X^i_t)$ for $i=1,\ldots,N$, with $v_t(X^i_t)$ given by \eqref{eq:kmed_particles_ode}\;
    $t\leftarrow t+\Delta t$\;
  }
  \KwOut{equal-weight particles $x^i_n := X^i_1$, $i=1,\ldots,N$, approximating $q_{n|n}$}
\end{\algoenv}

\paragraph{Discretisation.}
To obtain an implementable algorithm, we approximate $\rho_t$ by an empirical measure $\rho_t^N = \tfrac{1}{N}\sum_{i=1}^N \delta_{X_t^i}$, using $N\geq2$ particles, and discretise the differential equation \eqref{eq:kmed_ode} with the forward Euler method.
Specifically, the particle discretisation is achieved by estimating the likelihood expectation by its empirical mean:
\ifarxiv
\begin{equation}
  \hat r_t^i=\frac{p(y_n|X_t^i)}{N^{-1}\sum_{j=1}^{N}p(y_n|X_t^j)},
  \qquad \hat\sigma_t^2=\frac1{N-1}\sum_{i=1}^{N}(\hat r_t^i-1)^2,
  \qquad \hat h_t^i=-\frac{\hat r_t^i-1}{1+t\hat\sigma_t^2}.
  \label{eq:potential_particles}
\end{equation}
\else
\begin{align}
  \hat r_t^i&=\frac{p(y_n|X_t^i)}{N^{-1}\sum_{j=1}^{N}p(y_n|X_t^j)},
  &\hat\sigma_t^2&=\frac1{N-1}\sum_{i=1}^{N}(\hat r_t^i-1)^2,\nonumber\\
  \hat h_t^i&=-\frac{\hat r_t^i-1}{1+t\hat\sigma_t^2}.
  \label{eq:potential_particles}
\end{align}
\fi
Then \eqref{eq:kmed} yields the interacting particle system
\begin{subequations}
\label{eq:kmed_particles}
\begin{align}
  v_t(X_t^i)&=-\frac1N C_t\sum_{j=1}^N\alpha_t^j\nabla_{X_t^i}k(X_t^i,X_t^j),
  \label{eq:kmed_particles_ode}\\
  \boldsymbol h_t^k&=\left(\frac1N\mathbf G_t+\varepsilon I_N\right)\boldsymbol\alpha_t,
  \label{eq:kmed_particles_linear}
\end{align}
\end{subequations}
where $\boldsymbol\alpha_t=(\alpha_t^1,\ldots,\alpha_t^N)^\top$ and
\ifarxiv
\begin{equation}
  (\mathbf G_t)_{ij}=\frac1N\sum_{\ell = 1}^{N}\nabla_{X^\ell_t}k(X^i_t,X^\ell_t)\cdot C_t\nabla_{X^\ell_t}k(X^\ell_t,X^j_t),
  \qquad (\boldsymbol h_t^k)_i=\frac1{N-1}\sum_{j=1}^N k(X^i_t,X^j_t)\hat h_t^j.\label{eq:definition-linear-algebra}
\end{equation}
\else
\begin{align}
  &(\mathbf G_t)_{ij}=\frac1N\sum_{\ell = 1}^{N}\nabla_{X^\ell_t}k(X^i_t,X^\ell_t)\cdot C_t\nabla_{X^\ell_t}k(X^\ell_t,X^j_t), \notag\\
  &(\boldsymbol h_t^k)_i=\frac1{N-1}\sum_{j=1}^N k(X^i_t,X^j_t)\hat h_t^j.\label{eq:definition-linear-algebra}
\end{align}
\fi
Given a time step $\Delta t > 0$, the particle update from $X_t$ to $X_{t+\Delta t}$ consists of two steps. First, we solve the $N\times N$ linear system \eqref{eq:kmed_particles_linear} for $\boldsymbol\alpha_t$. Second, we compute the velocity field \eqref{eq:kmed_particles_ode} and update the particles by an Euler step,
\[
X^i_{t+\Delta t} = X^i_{t} + \Delta t \, v_t(X_t^i), \qquad i = 1,\ldots,N.
\]

\paragraph{Particle filtering with KMED.}
Given particles $\{x^i_{n-1}\}_{i=1}^N$ approximating the posterior $q_{n-1\mid n-1}$, the dynamical model \eqref{eq:dynamics} is simulated independently for each particle, $x^i_{n|n-1}=\mathcal M(x^i_{n-1},\epsilon_n^i)$, which gives an ensemble representation of the predictive distribution $q_{n\mid n-1}$. These particles are then transported from $t=0$ to $t=1$ by the KMED update above, which gives \Cref{alg:kmed_pro}. Particles remain equally weighted, and no transition density, resampling or likelihood gradient is needed.

\newcommand{\langevinFigure}[1]{%
\begin{figure*}[#1]
  \centering
  \includegraphics[width=\textwidth]{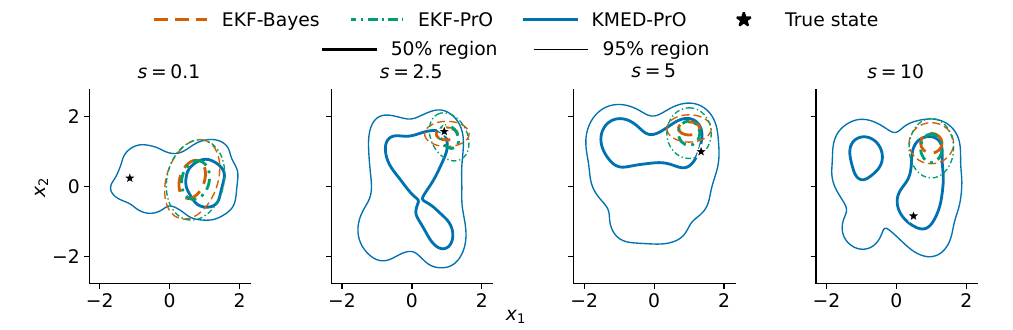}
  \caption{True hidden state trajectory and posterior approximations from each method at four checkpoints of a 100-observation run. EKF-Bayes and EKF-PrO concentrate on a single branch of the energy level set, and can miss the true state. KMED-PrO yields broader uncertainty regions that cover all branches consistent with the observed energy, and its $95\%$ regions always contain the true state.}
  \label{fig:2d-Langevin}
\end{figure*}}
\ifarxiv\else\langevinFigure{t}\fi

\subsection{Practical Considerations}
\label{subsec:practical}

As a preparation for \Cref{sec:experiments}, here we discuss practical considerations for implementation of KMED-PrO. Parameter values for each experiment are given in Appendix \ref{app:experiment-parameter}.

\paragraph{Preconditioning Matrix and Kernel Function.}
We choose the preconditioning matrix $C_t$ in \eqref{eq:ansatz} as the empirical covariance matrix for the ensemble at time $t\in[0,1]$
\begin{equation}
    \label{eq:empirical-covariance-preconditioning}
    C_t= \frac{1}{N-1} \sum_{i=1}^N(X_t^i-\bar X_t)(X_t^i-\bar X_t)^\top,
\end{equation}
where $\bar X_t=N^{-1}\sum_{i=1}^N X_t^i$.

For the kernel function, we employ an inverse multiquadric (IMQ) kernel
\begin{equation}
    \label{eq:IMQ-kernel}
    k_\lambda(x,y)=\bigl(1+\|x-y\|^2/\lambda^2\bigr)^{-1/2},
\end{equation}
with bandwidth $\lambda>0$, for all numerical experiments conducted in this paper. Unlike the Gaussian kernel, whose interactions vanish exponentially fast, the IMQ kernel has polynomially decaying tails, so it retains non-negligible interactions between distant particles \citep{gorham2017measuring, karvonen2026gaussian}. We choose the bandwidth adaptively with respect to the ensemble $\{X^i_t\}_{i=1}^N$ via the median heuristic \citep{gretton2012kernel, garreau2017large}, with a user-specified scaling factor. In some experiments, we instead fix the bandwidth for all $t\in[0,1]$.

\paragraph{Time Step.}
We use adaptive time steps over $t\in[0,1]$, with the step size depending on $\hat\sigma_t$ defined in \eqref{eq:potential_particles}:
\begin{equation}
  \Delta t=\min\left\{1-t,\;\delta\,\frac{1+t\hat\sigma_t^2}{\hat\sigma_t}\right\},
  \label{eq:adaptive_step}
\end{equation}
where $\delta\in(0,1)$ is a user-specified constant. Since $\hat h_t^i$ is the particle estimate of $h_t=-\partial_t\log\pi_t$ in \eqref{eq:pro_rate}, and
\[
  \Delta t\left(\frac1N\sum_i|\hat h_t^i|^2\right)^{1/2}
  \leq\sqrt{\frac{N-1}{N}}\,\delta\leq\delta,
\]
the rule \eqref{eq:adaptive_step} limits the first-order change in the log-density of the target path over each step to at most $\delta$, measured in $L^2(\rho_t^N)$, the empirical counterpart of the $L^2(\pi_t)$ norm in \Cref{prop:length}.
This is a heuristic based on the target path and does not by itself control the error of the particle approximation.
In particular, small steps are taken when the likelihood ratios vary substantially across the ensemble. The PrO path can be stiff close to $t=0$, since $\hat\sigma_0$ is large when $y_n$ is surprising under the predictive distribution, and \eqref{eq:adaptive_step} automatically takes small steps in this regime.
Moreover, since \eqref{eq:speed_bound} relies only on the inequality $1+t\sigma^2\geq 2\sqrt{t}\,\sigma$, it holds verbatim with $\hat\sigma_t$ in place of $\sigma_t$, and the number of steps taken by \eqref{eq:adaptive_step} is $O(\delta^{-1} + \log_+\log_+(\delta \hat\sigma_0))$, where $\log_+(z) = \max\{0,\log(z)\}$ (\Cref{app:steps}).
Because the $\hat r_t^i$ are non-negative with mean one, $\hat\sigma_0^2\leq N$, so the number of steps is bounded by $O(\delta^{-1}+\log_+\log_+(\delta\sqrt{N}))$ uniformly over the observation $y_n$.
By contrast, the analogous scheme for the Bayesian path \eqref{eq:bayes_path} takes a number of steps proportional to the Fisher--Rao length of that path, which grows without bound as $y_n$ becomes more outlying.

\paragraph{Computational Cost.}
Each time step requires $N$ likelihood evaluations, $O(N^2d^2+N^3d)$ operations to assemble the preconditioned Gram matrix and $O(N^3)$ operations to solve the linear system.
For $N > d$, the leading cost is $O(N^3d+Nc)$, with $c$ the cost of a likelihood evaluation.

\section{Empirical Results}
\label{sec:experiments}

In this section, we compare KMED-PrO with EKF-Bayes and EKF-PrO numerically. We consider two experiments. The first is a filtering task for the Langevin diffusion, where only the energy values of the hidden state are observed. The second is a range-tracking problem involving both partially observed data and a misspecified dynamical model.

\subsection{Langevin Filtering with an Energy Sensor}
\label{subsec:double-well}

In this section, we consider filtering for a Langevin diffusion \citep[Chapter 4]{pavliotis2014stochastic} whose target is a two-dimensional double-well distribution:
\begin{equation}
  \hspace{-5pt} \dd x_s^\star = -\nabla U(x_s^\star)\,\dd s + \dd W_s, \; { \ifarxiv \qquad \fi }
  y_n = U(x_{s_n}^\star) + 0.25\,\xi_n,
  \label{eq:Langevin-filter}
\end{equation}
where $U(x) = (x_1^2 - 1)^2 + x_2^2/2$ for $x = (x_1, x_2) \in \R^2$, $W_s$ is a standard two-dimensional Brownian motion, and the $\xi_n$ are independent standard Gaussian noises. We generate the true hidden trajectory starting from $(-1, 0)$ and observe it at times $s_n = 0.1n$ for $n = 1,\ldots,100$; see Appendix~\ref{app:langevin-details} for details of the simulation of the dynamical model. All filters use the same dynamical model and the same likelihood $p(y_n\mid x_n)=\mathcal N\big(y_n;U(x_n),0.25^2\big)$ as the data-generating process, together with the common initial prior $\mathcal{N}\big((0.8,\,0.2),\, \mathrm{diag}(1,0.16)\big)$.

\ifarxiv\langevinFigure{!htb}\fi

Figure \ref{fig:2d-Langevin} compares the true hidden state with the posterior approximations of each method at $n = 1, 25, 50, 100$. Both EKF variants (EKF-Bayes and EKF-PrO) perform poorly in this partially observed setting: they concentrate on a single branch whose energy is similar to that of the true state. EKF-PrO yields broader regions, but still excludes the true state at some of the displayed times. In contrast, KMED-PrO covers all regions with energy similar to that of the true state, and its $95\%$ regions contain the true state at all four displayed times. This illustrates the advantage of KMED-PrO in representing all states consistent with the observations when the state position is not directly observed.

\subsection{Partially Observed Range Tracking with Misspecified Dynamics}
\label{subsec:range-tracking}

In this section, we study a misspecified range-tracking problem and evaluate KMED-PrO as a function of the ensemble size $N$ and the dimension $d$, and check its sensitivity to the kernel bandwidth and to the regularisation of the linear system \eqref{eq:kmed_particles_linear}. We draw the initial hidden state $x_0^{\star}\in\R^d$ from $\mathcal N(0.5e_1, P_0)$, where $e_1=(1,0,\ldots,0)^\top$ and $P_0=\mathrm{diag}\big(1,\tfrac{1}{d-1},\ldots,\tfrac{1}{d-1}\big)$, and draw $S_0$ uniformly from $\{-1,1\}$. The true trajectory is then generated by
\[
  x_n^\star = 0.9\,x_{n-1}^\star + 0.15\,S_n e_1 + \epsilon_n^\star, \qquad n = 1, \ldots, 500,
\]
where $S_n$ flips the sign of $S_{n-1}$ with probability $0.04$, and the $\epsilon_n^\star$ are i.i.d.\ $\mathcal N(0,0.3^2\,P_0)$. Observations are
\[
  y_n = H_n(x_n^\star) + 0.3\,\xi_n, \qquad n = 1,\ldots,500,
\]
where the $\xi_n$ are independent standard Gaussian noises and, for $x=(x_1,\ldots,x_d)$,
\begin{equation}
  H_n(x)=
  \begin{cases}
    x_1, & 20 \mid n,\\[2pt]
    \sqrt{0.5^2+\|x\|^2}, & 20 \nmid n.
  \end{cases}
  \label{eq:observation-model-range-tracking}
\end{equation}
Thus, every $20$ steps only the first coordinate of the state is observed, and otherwise only a scalar range is available.

All filters start from the prior $\mathcal N(0.5e_1, P_0)$ and use the misspecified dynamical model $x_n = 0.9\,x_{n-1} + \epsilon_n$ with $\epsilon_n\sim\mathcal N(0,0.3^2P_0)$, together with the likelihood $\mathcal N(y_n; H_n(x_n), 0.3^2)$. For each experiment, we use $10$ random seeds and report the mean and standard deviation of the time-averaged negative log predictive density (NLPD), which measures how well a method predicts the next observation (lower is better); details are given in Appendix~\ref{app:range-details}.

\newcommand{\rangeDimensionCaption}{Time-averaged NLPD across dimensions at $N=128$ (mean $\pm$ SD over ten trajectories).}
\newcommand{\rangeParticleCaption}{Time-averaged NLPD across ensemble sizes at $d=10$ (mean $\pm$ SD over ten trajectories).}
\newcommand{\rangeSensitivityCaption}{Parameter sensitivity of KMED-PrO in range tracking ($d=5$, $N=128$). Each cell shows the mean time-averaged NLPD over ten random seeds for a given kernel bandwidth scaling factor and regularisation parameter $\varepsilon$ of the linear system \eqref{eq:kmed_particles_linear}. The outlined cell marks the setting used in \Cref{tab:range-dimension,tab:range-particles}; the hatched cell marks an unstable configuration.}
\newcommand{\rangeDimensionRows}{%
    \toprule
    $d$ & EKF-Bayes & EKF-PrO & KMED-PrO\\
    \midrule
    2 & $0.880 \pm 0.098$ & $0.933 \pm 0.120$ & $\mathbf{0.759 \pm 0.040}$\\
    25 & $0.793 \pm 0.087$ & $0.866 \pm 0.123$ & $\mathbf{0.678 \pm 0.060}$\\
    50 & $0.790 \pm 0.083$ & $0.838 \pm 0.074$ & $\mathbf{0.674 \pm 0.060}$\\
    \bottomrule
    }
\newcommand{\rangeParticleRows}{%
    \toprule
    & \multicolumn{2}{c}{Independent of $N$} & Particle-based\\
    \cmidrule(lr){2-3}\cmidrule(lr){4-4}
    $N$ & EKF-Bayes & EKF-PrO & KMED-PrO\\
    \midrule
    $32$  & $|$  & $|$ & $\mathbf{0.768 \pm 0.071}$\\
    $64$  & $0.802 \pm 0.077$ & $0.876 \pm 0.126$ & $\mathbf{0.693 \pm 0.069}$\\
    $128$ & $|$ & $|$ & $\mathbf{0.678 \pm 0.054}$\\
    \bottomrule
    }
\ifarxiv
\begin{figure}[b]
  \centering
  \begin{minipage}[c]{0.5\textwidth}
    \centering
    \footnotesize
    \setlength{\tabcolsep}{3pt}
    \captionof{table}{\rangeDimensionCaption}
    \label{tab:range-dimension}
    \begin{tabular}{@{}rccc@{}}
    \rangeDimensionRows
    \end{tabular}

    \vspace{12pt}
    \captionof{table}{\rangeParticleCaption}
    \label{tab:range-particles}
    \begin{tabular}{@{}rccc@{}}
    \rangeParticleRows
    \end{tabular}
  \end{minipage}\hfill
  \begin{minipage}[c]{0.46\textwidth}
    \centering
    \includegraphics[width=0.9\linewidth]{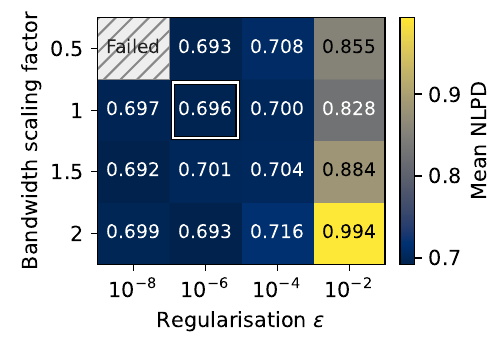}
    \caption{\rangeSensitivityCaption}
    \label{fig:range-sensitivity}
  \end{minipage}
\end{figure}
\else
\begin{table}[t]
  \centering
  \caption{\rangeDimensionCaption}
  \label{tab:range-dimension}
  \small
  \setlength{\tabcolsep}{3pt}
  \resizebox{\columnwidth}{!}{%
    \begin{tabular}{@{}rccc@{}}
    \rangeDimensionRows
    \end{tabular}%
  }
\end{table}

\begin{table}[t]
  \centering
  \caption{\rangeParticleCaption}
  \label{tab:range-particles}
  \small
  \setlength{\tabcolsep}{3pt}
  \begin{tabularx}{\linewidth}{@{}r*{3}{>{\centering\arraybackslash}X}@{}}
    \rangeParticleRows
  \end{tabularx}
\end{table}

\begin{figure}[!htbp]
  \centering
  \includegraphics[width=\columnwidth]{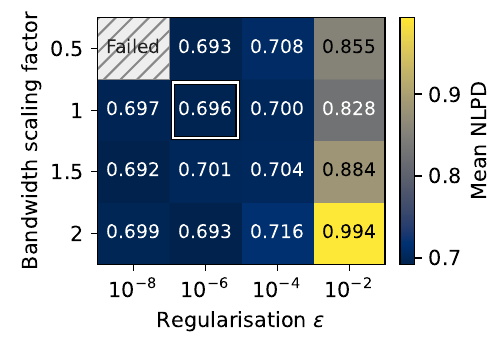}
  \caption{\rangeSensitivityCaption}
  \label{fig:range-sensitivity}
\end{figure}
\fi

Table \ref{tab:range-dimension} tests whether the predictive advantage of KMED-PrO persists as $d$ grows, with the total variance $\mathrm{tr}(P_0)=2$ held fixed. With $N=128$ particles, KMED-PrO attains the lowest time-averaged NLPD at every dimension, improving on EKF-Bayes by about $0.12$ per observation at $d=2,25,50$. The decrease of NLPD with $d$ for all methods is not informative about accuracy, since the distribution of the observations changes as $d$ grows. Table \ref{tab:range-particles} examines the effect of $N$ at $d=10$: KMED-PrO again attains the lowest time-averaged NLPD at all three tested values of $N$, and its NLPD falls as $N$ grows, with diminishing gains.

Finally, Figure \ref{fig:range-sensitivity} shows the sensitivity of KMED-PrO to the kernel bandwidth scaling factor and the regularisation parameter $\varepsilon$, evaluated over a grid of both parameters for $d=5$ and $N=128$. KMED-PrO is unstable when both the regularisation and the bandwidth are small. Large regularisation removes the instability but introduces substantial numerical error. This trade-off suggests using moderate values of both parameters.

\section{Related Work}
\label{sec:related}

\paragraph{Bayesian Filtering.}
Since the latent state $x_n$ may be only weakly constrained by the observations, Bayesian filtering is the dominant paradigm.
In the linear--Gaussian case the Bayesian filtering distribution is computed exactly by the Kalman filter \citep{kalman1960}, while the extended and unscented Kalman filters provide Gaussian approximations for nonlinear models \citep{Sarkka23}.
For nonlinear and non-Gaussian models, Monte Carlo methods are used.
Particle filters reweight and resample an ensemble of particles \citep{gordon1993novel,doucet2001sequential}, but their importance weights degenerate in all but low dimensions \citep{bengtsson2008curse,snyder2008obstacles}.
The ensemble Kalman filter (EnKF) avoids weights altogether, at the cost of a Gaussian assumption in the update \citep{evensen1994enkf,Evensen2009}.
A third class of methods instead \emph{transports} an unweighted ensemble from the predictive distribution to the filtering distribution, by integrating an ordinary differential equation in a pseudo-time variable \citep{daum2010exact,reich2011dynamical,bergemann2012ensemble,yang2013feedback,pulido2019sequential}.
Of particular relevance to this work are the \emph{kernel mean embedding dynamics} (KMED) of \citet{wang2024measure}, which construct this transport by matching the kernel mean embedding of the ensemble to that of a prescribed path of target distributions.
KMED requires only evaluations of the likelihood at the particles, and in the linear--Gaussian setting they recover the Kalman--Bucy flow when a quadratic kernel is used; see also the closely related \emph{kernel Fisher--Rao flow} of \citet{maurais2024sampling}.

\paragraph{Robust Filtering.}
The methods just described target the Bayesian filtering distribution, and therefore inherit its well-documented failure modes under model misspecification, which include overconfidence and divergence of the filter \citep{schlee1967divergencekf,fitzgerald2003divergencekf}.
Misspecification of the measurement model, for example due to outliers, has been addressed using heavy-tailed noise models \citep{Agamennoni2012,Huang2016}, robust losses \citep{boncelet1983,karlgaard2015nonlinear} and generalised Bayesian updates \citep{bissiri2016general,Knoblauch2022,boustati2020generalised,duran2024outlier,duran2025unifying}; these down-weight anomalous observations but cannot correct for systematic error.
Misspecification of the dynamical model has been addressed using heavy-tailed process noise \citep{roth2013studentfilter,huang2017ststfilter} and covariance inflation \citep{anderson1999enkf-covariance-inflation}, but these typically require prior knowledge of the form of misspecification, or additional hyperparameters to be tuned.

\paragraph{Predictively-Oriented Inference.}
PrO filtering \citep{shen2026predictively} builds on an emerging line of post-Bayesian research in which the expected log-likelihood in the variational characterisation of Bayes' rule is replaced by the log of the predictive likelihood \citep{masegosa2020learning,shenprediction,lai2024predictive,mclatchie2025predictively,liu2025detecting}.
Existing PrO filters are Gaussian, being based on the Kalman and extended Kalman filters \citep{shen2026predictively} or on the unscented transform \citep{kean2026unscented}; this paper provides the first nonparametric alternative.
Particle methods for PrO posteriors have been developed in the static setting \citep[e.g.][]{shenprediction,chazal2025computable,mclatchie2026markov}, but these require gradients and the score of the prior, which in filtering is the predictive distribution and is available only through samples.

\section{Limitations and Further Work}
\label{sec:discussion}

This paper introduced KMED-PrO, the first nonparametric, particle-based implementation of PrO filtering. 
It is enabled by a homotopy whose rate is available in closed form, even though the PrO filtering distribution is only implicitly defined. 
The illustrations show non-Gaussian hedging and retention of distinct state branches under partial observation. In range tracking with an omitted persistent force, KMED-PrO improves predictive NLPD over EKF-PrO and EKF-Bayes across the tested dimensions and particle counts. These comparisons assess the implemented filters; they do not isolate the PrO objective from the approximation family.
More broadly, our methods apply in principle to a broad class of implicitly defined variational targets, including batch PrO updates and static PrO posteriors whose prior is available only through samples.

Two limitations suggest further work. 
First, each pseudo-time step costs $O(N^3 d + N c)$, and the covariance preconditioner requires $N > d$.
Scaling to states whose nonlinear component is high-dimensional may therefore call for localisation or low-rank approximation of the Gram matrix. 
Second, Proposition \ref{prop:length} bounds the number of pseudo-time steps but not the approximation error. 
That error depends on the ensemble size, the kernel bandwidth, the regularisation $\varepsilon$ and the time discretisation, and it can propagate between assimilation steps. 
A large-$N$ convergence and stability analysis also remains open.

\section*{Acknowledgements}
C. J. Oates and L. Wang were supported by a Philip Leverhulme Prize (PLP-2023-004).
L. Wang and N. N\"usken acknowledge support from OpenAI's program for academic researchers.

\emph{AI statement.} The authors used large language models (ChatGPT and Claude) to assist with mathematical derivations and to improve the clarity of the writing, and used Codex to assist with implementing the algorithms. The research idea, the novel contributions, and the design of the numerical experiments are entirely the authors' own. All AI-assisted content, including derivations and code, was checked by the authors, who take full responsibility for the content of this paper.

\bibliographystyle{plainnat}
\bibliography{bibliography}

\appendix
\section*{Appendices}

Proofs for all theoretical results reported in the main text are contained in \Cref{app:proofs}.
Further details and additional results regarding the empirical assessment reported in \Cref{sec:experiments} are contained in \Cref{app:experiments}.

\section{Proofs}
\label{app:proofs}

This appendix contains the proofs of the results stated in \Cref{sec:methods}.
Throughout, $y_n$ is fixed, $p(y_n | \cdot)$ is abbreviated $p(\cdot)$ where no confusion can arise, and we assume that $p_{\pi_0}(y_n) > 0$.

\subsection{Proof of \texorpdfstring{\Cref{prop:rate}}{Proposition 2.1}}
\label{app:proof_rate}

\begin{proof}
	First we note that $r_t$ is bounded.
	Indeed, $p_{\pi_t}(y_n) > 0$, since by optimality the objective in \eqref{eq:pro_path} at $\pi_t$ is no larger than its value $- t \log p_{\pi_0}(y_n) < \infty$ at $q = \pi_0$, and $\sup_x p(y_n | x) < \infty$ by assumption.
	In particular $\sigma_t^2 < \infty$.

	Let $\dot{\ell}_t := \partial_t \log \pi_t$.
	Differentiating $\int \pi_t = 1$ gives $\E_{\pi_t}[\dot{\ell}_t] = 0$.
	Taking logarithms in $\pi_t \propto \pi_0 \exp(t r_t)$ and differentiating,
	\begin{align*}
		\dot{\ell}_t = r_t + t \, \partial_t r_t - \kappa_t ,
	\end{align*}
	where $\kappa_t$ is the derivative of the log-normalising constant, which does not depend on $x$.
	From \eqref{eq:ratio}, $\partial_t r_t = - r_t c_t$, where $c_t := \partial_t \log p_{\pi_t}(y_n)$, and hence
	\begin{align}
		\dot{\ell}_t = r_t (1 - t c_t) - \kappa_t . \label{eq:app_ell_dot}
	\end{align}
	On the other hand, differentiating $p_{\pi_t}(y_n) = \int p(y_n | x) \, \pi_t(x) \, \dd x$ and using $\E_{\pi_t}[\dot{\ell}_t] = 0$,
	\begin{align*}
		c_t = \frac{1}{p_{\pi_t}(y_n)} \int p(y_n | x) \, \dot{\ell}_t(x) \, \pi_t(x) \, \dd x = \E_{\pi_t}[r_t \dot{\ell}_t] = \Cov_{\pi_t}[r_t, \dot{\ell}_t] = (1 - t c_t) \, \sigma_t^2 ,
	\end{align*}
	where the final equality follows from \eqref{eq:app_ell_dot}.
	Solving this linear equation gives $c_t = \sigma_t^2 / (1 + t \sigma_t^2)$, which is the second claim, and $1 - t c_t = 1 / (1 + t \sigma_t^2)$.
	Substituting into \eqref{eq:app_ell_dot}, $\dot{\ell}_t = r_t / (1 + t \sigma_t^2) - \kappa_t$, and the constraint $\E_{\pi_t}[\dot{\ell}_t] = 0$ together with $\E_{\pi_t}[r_t] = 1$ gives $\kappa_t = 1 / (1 + t \sigma_t^2)$.
	Thus $\dot{\ell}_t = (r_t - 1) / (1 + t \sigma_t^2) = - h_t$, and since $\E_{\pi_t}[h_t] = 0$ this is \eqref{eq:pro_rate}.
\end{proof}

\subsection{Proof of \texorpdfstring{\Cref{prop:length}}{Proposition 2.3}}
\label{app:proof_length}

\begin{proof}
	From \Cref{prop:rate} and $\E_{\pi_t}[h_t] = 0$,
	\begin{align*}
		\| \partial_t \log \pi_t \|_{L^2(\pi_t)}^2 = \E_{\pi_t}[h_t^2] = \frac{\Var_{\pi_t}[r_t]}{(1 + t \sigma_t^2)^2} = \frac{\sigma_t^2}{(1 + t \sigma_t^2)^2} .
	\end{align*}
	The bound $\sigma_t / (1 + t \sigma_t^2) \leq \sigma_t$ is immediate, while for $t > 0$ the arithmetic--geometric mean inequality $1 + t \sigma_t^2 \geq 2 \sqrt{t} \, \sigma_t$ gives $\sigma_t / (1 + t \sigma_t^2) \leq 1 / (2 \sqrt{t})$, establishing \eqref{eq:speed_bound}.
	Integrating, $\int_0^1 \| \partial_t \log \pi_t \|_{L^2(\pi_t)} \, \dd t \leq \int_0^1 (2 \sqrt{t})^{-1} \, \dd t = 1$.

	For the second claim, fix $t \in [0,1]$.
	Recall from the proof of \Cref{prop:rate} that $p_{\pi_t}(y_n) > 0$, so that $0 \leq r_t \leq \sup_x p(y_n | x) / p_{\pi_t}(y_n) < \infty$.
	Since $\pi_t \propto \pi_0 \exp(t \, r_t)$, the distributions $\pi_t$ and $\pi_0$ are therefore mutually absolutely continuous, with $\log (\dd \pi_t / \dd \pi_0) = t \, r_t - \log Z_t$ bounded, where $Z_t := \E_{\pi_0}[\exp(t \, r_t)]$; in particular, both divergences in \eqref{eq:sym_kl} are finite.
	Taking the difference of the expectations of this log-density ratio under $\pi_t$ and $\pi_0$, the normalising constant cancels and
	\begin{align*}
		\KL(\pi_t \,\|\, \pi_0) + \KL(\pi_0 \,\|\, \pi_t) = \E_{\pi_t}\left[ \log \frac{\dd \pi_t}{\dd \pi_0} \right] - \E_{\pi_0}\left[ \log \frac{\dd \pi_t}{\dd \pi_0} \right] = t \left( \E_{\pi_t}[r_t] - \E_{\pi_0}[r_t] \right) .
	\end{align*}
	From \eqref{eq:ratio}, $\E_{\pi_t}[r_t] = 1$ and $\E_{\pi_0}[r_t] = p_{\pi_0}(y_n) / p_{\pi_t}(y_n) \geq 0$, which establishes \eqref{eq:sym_kl}.
	The final claim is the case $t = 1$, recalling that $\pi_0 = q_{n|n-1}$ and $\pi_1 = q_{n|n}$.
\end{proof}

\subsection{Number of Pseudo-Time Steps}
\label{app:steps}

Here we justify the claim in \Cref{subsec:practical} on the number of steps taken by the adaptive rule \eqref{eq:adaptive_step}.
Let $t_0 = 0$ and $t_{m+1} = t_m + \min\{1 - t_m, \delta / s_m\}$, where $s_m := \hat{\sigma}_{t_m} / (1 + t_m \hat{\sigma}_{t_m}^2)$ and $\delta \in (0, 1)$.
If $s_m=0$, the remaining interval is taken in one step; otherwise the divisions below are well-defined.
We use $\log_+(z)=\max\{0,\log z\}$ for $z>0$, with $\log_+(0)=0$.
Then $s_0 = \hat{\sigma}_0$, and the argument used to prove \eqref{eq:speed_bound} gives $s_m \leq 1 / (2 \sqrt{t_m})$ for $m \geq 1$, whatever the values of $\hat{\sigma}_{t_m}$.
Hence $t_1 \geq \min\{1, \delta / \hat{\sigma}_0\}$ and $t_{m+1} \geq \min\{1, t_m + 2 \delta \sqrt{t_m}\}$ for $m \geq 1$.

\paragraph{Phase 1 ($t_m < \delta^2$).}
Let $u_m := \log(\delta^2 / t_m) > 0$.
From $t_{m+1} \geq 2 \delta \sqrt{t_m}$ we obtain $u_{m+1} \leq \frac{1}{2} u_m - \log 2$, so $u_m$ at least halves at each step and becomes non-positive after at most $O(1 + \log_+ u_1)$ steps.
Since $u_1 \leq \log_+(\delta \hat{\sigma}_0)$, Phase 1 lasts for at most $O(1 + \log_+ \log_+ (\delta \hat{\sigma}_0))$ steps.

\paragraph{Phase 2 ($t_m \geq \delta^2$).}
Writing $\tau_m := \sqrt{t_m} \geq \delta$, we have $\tau_{m+1} \geq \sqrt{\tau_m^2 + 2 \delta \tau_m}$, and since $2 \delta \tau_m \leq 2 \tau_m^2$,
\begin{align*}
	\tau_{m+1} - \tau_m \geq \frac{2 \delta \tau_m}{\sqrt{\tau_m^2 + 2 \delta \tau_m} + \tau_m} \geq \frac{2 \delta}{1 + \sqrt{3}} ,
\end{align*}
unless $t_{m+1} = 1$.
Thus Phase 2 lasts for at most $\lceil (1 + \sqrt{3}) / (2 \delta) \rceil + 1$ steps.

Combining the two phases, the total number of steps is at most $C\,(\delta^{-1} + \log_+ \log_+(\delta \hat{\sigma}_0))$ for a numerical constant $C$ that does not depend on $\delta$, $y_n$ or the particles.
Finally, since $\hat r_0^i\geq 0$ and $\frac1N\sum_i\hat r_0^i=1$, we have $\sum_i(\hat r_0^i)^2\leq N^2$ and hence $\hat\sigma_0^2=\frac{1}{N-1}\bigl(\sum_i(\hat r_0^i)^2-N\bigr)\leq N$, so the bound above is uniform in $y_n$.

\section{Numerical Experiments}
\label{app:experiments}

\subsection{Computational Environment}
All numerical results were generated on a local MacBook Pro with an Apple M5 chip (10 CPU cores, 32 GB of memory), running macOS 26.6.2. All experiments were implemented in \emph{PyTorch} (version 2.14.1) with Python 3.12.13, using CPU execution and double-precision (\texttt{float64}) arithmetic. Experiment-specific random seeds were fixed in advance and recorded in the corresponding configurations for reproducibility.

\subsection{Algorithm Setup}
\label{app:experiment-parameter}

We consider the Gaussian-noise model
\[
\begin{aligned}
    x_n &= f_n(x_{n-1}) + \epsilon_n,
    & \epsilon_n &\sim \mathcal N(0,Q_n), \\
    y_n &= g_n(x_n) + \xi_n,
    & \xi_n &\sim \mathcal N(0,R_n),
\end{aligned}
\]
where $x_n\in\mathbb R^d$, $y_n\in\mathbb R^p$, and the noise
variables are mutually independent and independent of the initial state.

For both EKF variants, given the previous posterior approximation
$\mathcal N(m_{n-1},P_{n-1})$, the Gaussian forecast
$\mathcal N(a_n,B_n)$ is obtained by linearising the dynamical map:
\[
    F_n = Df_n(m_{n-1}), \qquad
    a_n = f_n(m_{n-1}), \qquad
    B_n = F_nP_{n-1}F_n^\top+Q_n,
\]
where $D$ denotes the Jacobian with respect to the state variable.

\paragraph{EKF-Bayes.}
We use the affine linearisation of the measurement map at the forecast mean,
\[
    G_n^{\mathrm B}(x)=g_n(a_n)+H_n(x-a_n),
    \qquad H_n=Dg_n(a_n).
\]
The updated Gaussian $\mathcal N(m_n,P_n)$ is obtained from
\[
\begin{gathered}
    K_n=B_nH_n^\top(H_nB_nH_n^\top+R_n)^{-1},
    \qquad m_n=a_n+K_n\bigl(y_n-G_n^{\mathrm B}(a_n)\bigr), \\
    P_n=(I-K_nH_n)B_n(I-K_nH_n)^\top+K_nR_nK_n^\top.
\end{gathered}
\]

\paragraph{EKF-PrO.}
Following \citet{shen2026predictively}, we initialise
$m_n^{(0)}=a_n$ and $P_n^{(0)}=B_n$, keeping $a_n,B_n$ fixed throughout
the update. At outer iteration $j$, we linearise the same measurement
map $g_n$ at the current mean and define
\[
\begin{gathered}
    G_n^{(j)}(x)=g_n(m_n^{(j-1)})+H_j(x-m_n^{(j-1)}),
    \qquad H_j=Dg_n(m_n^{(j-1)}), \\
    v_j=y_n-G_n^{(j)}(a_n),\quad
    S_j(P)=H_jPH_j^\top+R_n,\quad
    M_j(P)=S_j(P)+H_jB_nH_j^\top.
\end{gathered}
\]
With $G_n^{(j)}$ fixed, we approximately minimise the convex covariance
surrogate
\[
    \varphi_j(P)=\frac12\Bigl[
        \operatorname{tr}(B_n^{-1}P)-\log|P|
        +v_j^\top M_j(P)^{-1}v_j
        +\operatorname{tr}\!\left(S_j(P_n^{(j-1)})^{-1}S_j(P)\right)
    \Bigr].
\]
We use the limited-memory Broyden--Fletcher--Goldfarb--Shanno (L-BFGS)
algorithm \citep[Algorithm 2.1]{liu1989limited}, which approximates curvature from recent iterates and gradients.
Initialising the inner solve at $P_n^{(j-1)}$, we compute
\[
    P_n^{(j)}=\operatorname*{arg\,min}_{P\succ0}\varphi_j(P),
    \qquad m_n^{(j)}=a_n+B_nH_j^\top M_j(P_n^{(j)})^{-1}v_j.
\]
We then relinearise at $m_n^{(j)}$. After the fixed number $J$ of outer
iterations specified in Table \ref{tab:numerical-settings}, we set
$(m_n,P_n)=(m_n^{(J)},P_n^{(J)})$.

Each inner solve uses a positive-definite Cholesky parameterisation,
L-BFGS memory size $20$, and budgets of $100$ iterations and $3000$
L-BFGS objective evaluations. It stops when
\[
    \|\nabla_P\varphi_j(P_{\mathrm{new}})\|_{\mathrm F}\le10^{-4}
    \quad\text{or}\quad
    \frac{\varphi_j(P_{\mathrm{old}})-\varphi_j(P_{\mathrm{new}})}
         {\max\{1,\varphi_j(P_{\mathrm{old}})\}}\le10^{-7},
\]
where the second criterion must hold for three consecutive inner iterations.
Here $\|\cdot\|_{\mathrm F}$ is the Frobenius norm, and
$P_{\mathrm{old}},P_{\mathrm{new}}$ are successive inner iterates.
Budget exhaustion before either stopping criterion is met is reported
as failure.

\paragraph{KMED-PrO.}
Instead of a Gaussian approximation, KMED-PrO represents the posterior at filtering step $n-1$ by an ensemble of samples $\{x^i_{n-1}\}_{i=1}^{N}$. The ensemble representing the posterior at step $n$ is then computed by Algorithm \ref{alg:kmed_pro}.

The parameters of each method are listed for each experiment in Table \ref{tab:numerical-settings}.

\begin{table}[htbp]
  \centering
  \small
  \setlength{\tabcolsep}{4pt}
  \caption{Numerical settings for the displayed results.}
  \label{tab:numerical-settings}
  \begin{tabularx}{\linewidth}{@{}Xrrrrlr@{}}
  \toprule
  Experiment & $d$ & $N$ & $\delta$ & $\varepsilon$ & Bandwidth & $J$\\
  \midrule
  Direct-position update (Fig \ref{fig:teaser}(a)) & 1 & 256 & 0.01 & $10^{-8}$ & median heuristic & 20\\
  Static squared sensor (Fig \ref{fig:teaser}(b)) & 1 & 256 & 0.0025 & $10^{-3}$ & fixed, $2$ & 20\\
  2d-Langevin (Fig \ref{fig:2d-Langevin}) & 2 & 256 & 0.01 & $10^{-5}$ & $0.25\times$ median heuristic & 20\\
  Range tracking, varying $d$ (Table \ref{tab:range-dimension}) & $2,25,50$ & 128 & 0.01 & $10^{-6}$ & median heuristic & 30\\
  Range tracking, varying $N$ (Table \ref{tab:range-particles}) & 10 & $32,64,128$ & 0.01 & $10^{-6}$ & median heuristic & 30\\
  \bottomrule
  \end{tabularx}
\end{table}

\subsection{Introductory Illustrations}
\label{app:scalar-experiments}

\paragraph{Reference PrO posterior.}
The reference densities for the PrO posteriors in Figure~\ref{fig:teaser} are computed numerically, as described below.

\emph{Fixed-point formulation.} Let $\pi_0$ denote the prior density and $y$ the observation. The tempered PrO posterior at $t\in[0,1]$ is defined implicitly by
\[
\pi_t(x) = \frac{\pi_0(x)\exp\!\left(t\,\dfrac{p(y\mid x)}{b_t}\right)}{Z_t},
\]
where
\[
b_t := \int_{\R} p(y\mid x)\,\pi_t(x)\,\mathrm{d}x,
\qquad
Z_t := \int_{\R}\pi_0(x)\exp\!\left(t\,\frac{p(y\mid x)}{b_t}\right)\mathrm{d}x .
\]
Once $b_t$ is known, $\pi_t$ is determined. To find $b_t$, define, for $b>0$,
\[
g_t(b) := \int_{\R} p(y\mid x)\,\frac{\pi_0(x)\exp\!\left(t\,p(y\mid x)/b\right)}{Z_t(b)}\,\mathrm{d}x,
\qquad
Z_t(b) := \int_{\R}\pi_0(x)\exp\!\left(t\,\frac{p(y\mid x)}{b}\right)\mathrm{d}x,
\]
so that $b_t$ solves $g_t(b)=b$. Differentiating gives
\[
g_t'(b) = -\frac{t}{b^2}\,\mathrm{Var}_{\pi_t}\!\big(p(y\mid\cdot)\big)\le 0,
\]
so $g_t(b)-b$ is strictly decreasing in $b$ and the equation has at most one root.

\emph{Numerical scheme.} We approximate the PrO path at the times $t_j=j/50$, $j=0,\ldots,50$. At each $t_j$, we solve $g_{t_j}(b)=b$ by bisection on the scale $\log b$. The integrals defining $Z_{t_j}(b)$ and $g_{t_j}(b)$ are evaluated with the composite trapezoidal rule on $10241$ equally spaced points over $[-10,10]$. Substituting the resulting root $b_{t_j}$ into the definition of $\pi_t$ gives $\pi_{t_j}$, and the endpoint $\pi_1$ is the reference PrO posterior.

Both introductory illustrations below use random seed $1000$.

\paragraph{Direct-position update.}
Figure \ref{fig:teaser}(a) uses the forecast $q_{n|n-1}=\mathcal N(0,1)$, the likelihood $p(y\mid x)=\mathcal N(y;x,0.5^2)$, and the observation $y=3$. The Bayesian posterior is then $\mathcal N(2.4,0.2)$. The background shows the reference density path of the PrO posterior (see the paragraph above), together with $64$ recorded KMED-PrO particle trajectories. At the endpoint, the KMED-PrO density is a Gaussian kernel density estimate built from all $256$ transported particles, with bandwidth $0.2961$ estimated from the initial ensemble. The Gaussian densities of the EKF variants are plotted directly from their closed-form means and variances.

\paragraph{Static filtering with squared sensor.}
Figure \ref{fig:teaser}(b) uses identity dynamics without process noise and the prior $\mathcal N(0,4)$. The likelihood is $p(y_n\mid x_n)=\mathcal N(y_n;x_n^2,0.5^2)$, and the $20$ observations are generated as $y_n=4+0.5\,\xi_n$ with independent standard normal $\xi_n$. The figure displays the posterior distributions at $n=20$, after assimilating all $20$ observations.

Since the Gaussian mean is zero, the Jacobian of the sensor $x\mapsto x^2$ vanishes there, so the EKF-PrO update is uninformative and the estimate stays at the prior throughout this example. The displayed KMED-PrO density is a Gaussian kernel density estimate built from all $256$ final particles, with bandwidth $0.12$. 

\subsection{Langevin Filtering}
\label{app:langevin-details}

\paragraph{Dynamical Model Simulation.}
Instead of doing an Euler--Maruyama discretisation for the SDE in \eqref{eq:Langevin-filter}, we use an exact drift for the first coordinate followed by an independent Brownian increment for the first coordinate, and an exact Ornstein--Uhlenbeck transition for the second coordinate. 

Specifically, given a physical time step with length $h=0.01$, we compute
\[
x_{s+h,1} = \frac{x_{s,1}}{\sqrt{e^{-8h}+(1-e^{-8h})x_{s,1}^2}} + \sqrt{h}\epsilon_{s,1}, \qquad x_{s+h,2} = e^{-h}x_{s,2} + \sqrt{\frac{1 - e^{-2h}}{2}} \epsilon_{s,2}.
\]
where $\epsilon_{s,1}, \epsilon_{s,2}$ are independent i.i.d. 1-dimensional standard Gaussian noise. The displayed trajectory in Figure \ref{fig:2d-Langevin} uses random seed $1000$, with $100$ observations and checkpoints $s=0.1,2.5,5,10$. 

\paragraph{Density estimation.}
\emph{EKF variants.} The posterior of an EKF variant is Gaussian, $\mathcal{N}(x;m,P)$ with $m\in\R^2$ and $P\in\R^{2\times 2}$. For $c\in(0,1)$, its $c$-credible region is the ellipse
\[
R_c := \left\{x\in\R^2: (x - m)^\top P^{-1}(x-m) \leq -2\log(1-c)\right\},
\]
centered at $m$, which satisfies $\int_{R_c}\mathcal{N}(x;m,P)\,\mathrm{d}x = c$.

\emph{KMED-PrO.} Given the posterior particles $\{x^i\}_{i=1}^N\subset\R^2$, we first form the Gaussian kernel density estimate
\[
\hat q(x) = \frac{1}{N}\sum_{i=1}^N \frac{1}{2\pi\sqrt{\det B}}\exp\left(-\frac{1}{2}(x - x^i)^\top B^{-1} (x-x^i)\right), \qquad B = N^{-1/3}\,\hat C,
\]
where $\hat C$ is the empirical covariance of the posterior ensemble. This bandwidth follows Scott's rule \citep[Chapter~6]{scott1992multivariate}. For a threshold $\lambda>0$, let $R_\lambda := \{x\in\R^2: \hat q(x) \geq \lambda\}$. We seek the threshold $\lambda_c$ such that $\int_{R_{\lambda_c}} \hat q(x)\,\mathrm{d}x = c$, and approximate it numerically on a grid as follows.
\begin{enumerate}
  \item \emph{Grid.} Let $x^i_j$ denote the $j$-th coordinate of $x^i$. We take the rectangle
  \[
  \mathcal{D} = \prod_{j=1}^{2} \left[\min_i x^i_j - 7.5\sqrt{B_{jj}},\; \max_i x^i_j + 7.5\sqrt{B_{jj}}\right]
  \]
  and divide it uniformly into $241\times 241$ cells, each of area $\Delta A$. Let $K=241^2$ be the number of cells.
  \item \emph{Ordering.} Evaluate $\hat q$ at each cell center and sort the values in decreasing order, $d_1\geq d_2\geq\cdots\geq d_K$.
  \item \emph{Cumulative mass.} Define $S_0=0$ and $S_k = \Delta A\sum_{j=1}^{k} d_j$. Find the index $k$ with $S_{k-1} < c \leq S_k$.
  \item \emph{Threshold.} Set $\lambda_c = (1-f)\,d_{k-1} + f\,d_k$, where $f = \dfrac{c - S_{k-1}}{d_k\,\Delta A}\in(0,1]$.
\end{enumerate}
The $c$-confident region is then $R_{\lambda_c}$.

\subsection{Range Tracking}
\label{app:range-details}

\paragraph{Random seeds and experimental design.}
Tables \ref{tab:range-dimension} and \ref{tab:range-particles} are based on ten independent trajectories of $500$ observations each, generated with seeds $200101$--$200110$.

In Table \ref{tab:range-particles}, each seed uses the same observation trajectory for all particle counts. The initial particle draws and forecast innovations are also shared across ensemble sizes. The EKF variants do not depend on $N$, so their results are computed once and reported only once in the table.

The sensitivity study of KMED-PrO in Figure \ref{fig:range-sensitivity} uses a separate set of seeds, $210101$--$210110$, with $d=5$ and $N=128$. The same ten seeds, and hence the same observation trajectories, are used at every grid point. We consider bandwidth scaling factors in $\{0.5,\,1,\,1.5,\,2\}$ and regularisation parameters $\varepsilon\in\{10^{-8},\,10^{-6},\,10^{-4},\,10^{-2}\}$.

\paragraph{Time-averaged NLPD.}
At filtering step $n$, the predictive density of the observation is obtained from the forecast distribution $q_{n|n-1}$:
\begin{equation}
    p_n(y)=\int\mathcal N\!\left(y;H_n(x),\sigma^2\right)\, q_{n|n-1}(\dd x),
\label{eq:range-nlpd}
\end{equation}
where $H_n$ is the measurement model \eqref{eq:observation-model-range-tracking} and $\sigma=0.3$ is the observation noise standard deviation.

\emph{KMED-PrO.} Given the forecast particles $\{x_{n|n-1}^i\}_{i=1}^N\subset\R^d$, we approximate \eqref{eq:range-nlpd} by the Gaussian mixture
\[
 \widehat p_n(y)=\frac1N\sum_{i=1}^N
 \mathcal N\!\left(y;H_n(x_{n|n-1}^i),\sigma^2\right).
\]

\emph{EKF variants.} Let $\mathcal N(m_n,P_n)$ denote the forecast distribution. At the observation indices $n=20,40,\ldots,500$, the sensor is linear, $H_n(x)=x_1$, so the predictive density is available in closed form:
\[
\widehat p_n(y)
=
\mathcal N\!\left(y;m_{n,1},P_{n,11}+\sigma^2\right),
\]
where the subscript $1$ denotes the first coordinate. At the remaining indices, we approximate \eqref{eq:range-nlpd} numerically by Gauss--Hermite and generalised Gauss--Laguerre quadrature, as implemented in SciPy.

\emph{Metric.} For each seed, the time-averaged NLPD is $ -\frac{1}{500}\sum_{n=1}^{500}\log p_n(y_n)$, where $y_n$ is the observation at step $n$. We report the mean and standard deviation over the $10$ seeds.

\end{document}